\documentstyle[preprint,aps]{revtex}

\begin{document}
\scrollmode
\draft
\title{Exotic black hole solution in teleparallel theory of
(2+1)-dimensional gravity}
\author{Toshiharu Kawai\thanks{
Electronic address: h1758@ocugw.cc.osaka-cu.ac.jp}}
\address{Department of Physics, Osaka City University, 3-3-138
Sugimoto, Sumiyoshi-ku, Osaka 558, Japan}
\maketitle
\begin{abstract}
A black hole solution in a teleparallel theory of (2+1)-dimensional
gravity, given in a previous paper, is examined. This solution is also
a solution of the three-dimensional vacuum Einstein equation with
a vanishing cosmological constant. Remarkable is the fact that this
solution gives a black hole in a \lq \lq flat-land" in the Einstein
theory and a Newtonian limit. Coordinate transformations to
\lq \lq Minkowskian" coordinates, however, are
singular not only at the origin, but also on the event horizon.
{\em In the three-dimensional Einstein theory, vacuum regions of
space-times can be locally non-trivial}.
\end{abstract}

\narrowtext
\section{INTRODUCTION}
                            \setcounter{equation}{0}
In a previous paper [1], we have developed a teleparallel theory of
(2+1)-dimensional gravity, in which several static circularly
symmetric exact solutions of the gravitational field equation in
vacuum are given. The solution discussed there mainly is the solution
for the case with $3\alpha \neq -4\beta $, and with
$\alpha \beta \neq 0$, and two solutions for the case with
$3\alpha = -4\beta \neq 0$ have been mentioned in the APPENDIX B only
briefly.

The main purpose of this paper is to examine one of these two
solutions which gives a black hole in a \lq \lq flat-land" in the
three-dimensional Einstein theory and a Newtonian limit.
\section{BASIC FRAMEWORK OF TELEPARALLEL THEORY OF (2+1)-DIMENSIONAL
GRAVITY}
                                       \setcounter{equation}{0}
For the convenience of the latter discussion, we briefly summarize the
basic part of the teleparallel theory developed in Ref. [1].

The three-dimensional space-time is assumed to be a differentiable
manifold endowed with a Lorentzian metric
$g_{\mu \nu }dx^{\mu }\otimes dx^{\nu }$ related to the fields
${\bf e}^{k}={e^{k}}_{\mu }dx^{\mu }\; \; (k=0, 1, 2)$ through the
relation $g_{\mu \nu }={e^{k}}_{\mu }\eta_{kl}{e^{l}}_{\nu }$ with
$(\eta_{kl}) \stackrel{\rm def}{=}$diag$(-1, 1, 1)$. Here,
$\{x^{\mu };\mu=0, 1, 2\}$ is a local coordinate of the space-time.
The fields ${\bf e}_{k}={e^{\mu }}_{k}\partial /\partial x^{\mu }$,
which are dual to ${\bf e}^{k}$ are the dreibein fields.
The strength of ${e^{k}}_{\mu }$ is given by
\begin{equation}
    {T^{k}}_{\mu \nu }=\partial_{\mu }{e^{k}}_{\nu }
    -\partial_{\nu }{e^{k}}_{\mu }\;.
\end{equation}
The covariant derivative of the Lorentzian vector field $V^{k}$ is
defined
by
\begin{equation}
\nabla_{l}{\bf V}^{k}={e^{\mu }}_{l}\partial_{\mu }{\bf V}^{k}\;,
\end{equation}
and the covariant derivative of the world vector field ${\bf V}=
V^{\mu }\partial /\partial x^{\mu }$ with respect to the affine
connection $\Gamma^{\mu }_{\lambda \nu }$ is given by
\begin{equation}
\nabla_{\nu }V^{\mu }=\partial_{\nu }V^{\mu }
+\Gamma^{\mu}_{\lambda \nu }V^{\lambda }\;.
\end{equation}
The requirement
\begin{equation}
\nabla_{l}V^{k}={e^{\nu }}_{l}{e^{k}}_{\mu }\nabla_{\nu }V^{\mu }
\end{equation}
for $V^{\mu } \stackrel{\rm def}{=}{e^{\mu }}_{k}V^{k}$ leads to
\begin{eqnarray}
{T^{k}}_{\mu \nu }
&\equiv &{e^{k}}_{\lambda }{T^{\lambda }}_{\mu \nu }\;,\\
{R^{\mu }}_{\nu \lambda \rho } &\stackrel{\rm def}{=}&
\partial_{\lambda }\Gamma^{\mu }_{\nu \rho }-
\partial_{\rho }\Gamma^{\mu }_{\nu \lambda }+
\Gamma^{\mu }_{\tau \lambda }\Gamma^{\tau }_{\nu \rho }-
\Gamma^{\mu }_{\tau \rho }\Gamma^{\tau }_{\nu \lambda }\equiv 0\;,\\
\nabla_{\lambda }g_{\mu \nu } &\stackrel{\rm def}{=}&
\partial_{\lambda }g_{\mu \nu }-
\Gamma^{\rho }_{\mu \lambda }g_{\rho \nu }-
\Gamma^{\rho }_{\nu \lambda }g_{\rho \mu }\equiv 0\;,
\end{eqnarray}
where ${T^{\lambda }}_{\mu \nu }$ is defined by
\begin{equation}
{T^{\lambda }}_{\mu \nu }\stackrel{\rm def}{=}
\Gamma^{\lambda }_{\nu \mu }-\Gamma^{\lambda }_{\mu \nu }\;.
\end{equation}
The components ${T^{\lambda }}_{\mu \nu }$ and
${R^{\mu }}_{\nu \lambda \rho }$ are those of the torsion tensor and
of the curvature tensor, respectively. Equation (2.6) implies the
teleparallelism. The field components ${e^{k}}_{\mu }$ and
${e^{\mu }}_{k}$ are used to convert Latin and Greek indices. Also,
raising and lowering the indices $k, l, m, ...$ are accomplished
with the aid of $(\eta^{kl})=(\eta_{kl})^{-1}$ and $(\eta_{kl})$,
respectively.

For the matter field $\varphi $ belonging to a representation of the
three-dimensional Lorentz group,
$\L_{M}(\varphi , \nabla_{k}\varphi )$ with $\nabla_{k}\varphi
\stackrel{\rm def}{=}{e^{\mu }}_{k}\partial_{\mu }\varphi $ is a
Lagrangian density invariant under global Lorentz transformation
and under general coordinate transformation, if
$L_{M}(\varphi ,\partial_{k}\varphi )$ is an invariant Lagrangian
density on the three-dimensional Minkowski space-time. For the
dreibein fields ${\bf e}_{k}$, we have employed [1]
\begin{equation}
L_{G}=\alpha t^{klm}t_{klm}+\beta v^{k}v_{k}+\gamma a^{klm}a_{klm}\;,
\end{equation}
as the Lagrangian density. Here, $t_{klm}, v_{k}, $ and $a_{klm}$
are the irreducible components of $T_{klm}$, which are defined by
\begin{eqnarray}
 t_{klm} & \stackrel{\rm def}{=} & \frac{1}{2}(T_{klm}+T_{lkm})
 +\frac{1}{4}(\eta_{mk}v_{l}+\eta_{ml}v_{k})
 -\frac{1}{2}\eta_{kl}v_{m}\;, \nonumber \\
     & & \\
 v_{k} & \stackrel{\rm def}{=} & {T^{l}}_{lk}\;,
\end{eqnarray}
and
\begin{equation}
a_{klm}\stackrel{\rm def}{=}\frac{1}{3}(T_{klm}+T_{mkl}+T_{lmk})\;,
\end{equation}
respectively, and $\alpha ,\beta $, and $\gamma $ are real constant
parameters.  Then,
\begin{equation}
{\bf I}\stackrel{\rm def}{=}\frac{1}{c}\int {\bf L}d^{3}x
\end{equation}
is the total action of the system, where $c$ is the light velocity
in vacuum and ${\bf L}$ is defined by
\begin{equation}
{\bf L} \stackrel{\rm def}{=} \sqrt{-g}[L_{G}+
L_{M}(\varphi , \nabla_{k}\varphi )]
\end{equation}
with $g \stackrel{\rm def}{=} \det(g_{\mu \nu })$.
\section{STATIC CIRCULARLY SYMMETRIC SOLUTIONS OF GRAVITATIONAL
FIELD EQUATION IN VACUUM FOR THE CASE WITH $3\alpha =-4\beta \neq 0$}
                                        \setcounter{equation}{0}
For static circularly symmetric gravitational field,
$({e^{k}}_{\mu })$ can be assumed, without loss of generality, to
have a diagonal form:
\begin{eqnarray}
({e^{k}}_{\mu })=\left[\begin{array}{ccc} A(r) & 0 & 0 \\
                 0 & B(r) & 0 \\
                 0 & 0 & B(r) \end{array}\right]\;
\end{eqnarray}
with $A$ and $B$ being functions of $r \stackrel {\rm def}{=}
\sqrt{(x^{1})^{2}+(x^{2})^{2}}$, which leads to $a_{klm}\equiv 0$.

In Ref. [1], exact solutions of the field equation in vacuum for
$({e^{k}}_{\mu })$ having the expression (3.1) have been given for
the case with $3\alpha \neq -4\beta $ and
with $\alpha \beta \neq 0$ and for the case with
$3\alpha =-4\beta \neq 0$, and the former case has been discussed
in detail. For the latter case, we have obtained [1] two solutions:
\begin{equation}
A(r)=1+a\ln \left(\frac{r}{r_{0}}\right)\;, \; \; \;
B(r)=\frac{r_{0}}{r}\;,
\end{equation}
and
\begin{equation}
A(r)=1\;,\; \; \; B(r)=\left(\frac{r_{0}}{r}\right)^{b}\;,
\end{equation}
where $a$ and $b$ are constants and we have normalized as
$A(r_{0})=1=B(r_{0})$ for a positive $r_{0}$.
Since $3\alpha =-4\beta \neq 0$ and the tensor $a_{klm}$ identically
vanishes for $({e^{k}}_{\mu })$ having the expression (3.1), the
solutions (3.2) and (3.3) are also solutions of the three-dimensional
vacuum Einstein equation without cosmological term. This is known from
($3.7^{\prime }$) of Ref. [1].

We examine first the solution (3.2).

For a macroscopic test body such that the effects due to the intrinsic
\lq \lq spin" [2] of fundamental constituent particles can be ignored,
the equation of motion agrees with the geodesic line of the metric
$g_{\mu \nu }dx^{\mu }\otimes dx^{\nu }$ [1]. For the solution
(3.2), the equation reduces to
\begin{equation}
\frac{d^{2}x^{\alpha }}{dt^{2}}=-\left(1+a\ln \frac{r}{r_{0}}\right)
\frac{ac^{2}}{{r_{0}}^{2}}x^{\alpha }\;, \; \alpha =1\;, 2\;, \;
\mbox{for $r \simeq r_{0}$}\;,
\end{equation}
when the motion is sufficiently slow and terms quadratic in
$dx^{\alpha }/dt$ can be ignored. Here,
$t \stackrel{\rm def}{=} x^{0}/c$ is the coordinate time. On the
circle $r=r_{0}$, this agrees with the Newton equation of motion
\begin{equation}
\frac{d^{2}x^{\alpha }}{dt^{2}}=-2GM\frac{x^{\alpha }}{r^{2}}\;,
\end{equation}
for a particle moving under the influence of the Newton gravitatioal
force due to a point particle being located at the origin $r=0$
and having a mass $M$, if
\begin{equation}
ac^{2}=2GM\;.
\end{equation}
Also, around $r=r_{0}$, the potential
\begin{equation}
U \stackrel{\rm def}{=}-\frac{c^{2}}{2}(g_{00}-\eta_{00})
=ac^{2}\ln \frac{r}{r_{0}}+\frac{a^{2}c^{2}}{2}
\left(\ln \frac{r}{r_{0}}\right)^{2}
\end{equation}
is approximately equal to the Newton potential $2GM\ln (r/r_{0})$,
if the condition (3.6) is satisfied. Thus, the solution (3.2) gives a
Newtonian limit in the same sense as in the case of a theory [3] of
(1+1)-dimensional gravity.

The function $A$ has a zero for positive $r$, which shall be denoted
by $l$, i.e. $A(l)=0$. As has been mentioned in Ref. [1], the circle
$r=l$ is an event horizon, and the solution (3.2) gives a black hole
space-time. Also, (3.2) {\em gives a black hole in a
\lq \lq flat-land" in the Einstein theory}, because it is also a
solution of the three-dimensional vacuum Einstein equation and the
Riemann-Christoffel curvature tensor [4] vanishes for $r\neq 0$.

There is a coordinate system $\{y^{k};k=0,1,2\}$ for
which the metric tensor takes the Minkowskian form
\begin{equation}
\eta_{kl}dy^{k}\otimes dy^{l}=
g_{\mu \nu }dx^{\mu }\otimes dx^{\nu }\;,\; \; \; {\rm for}\;\;\;
r\neq 0\;.
\end{equation}
The coordinate system $\{y^{k};k=0,1,2\}$ is given by a solution
of the equations,
\begin{eqnarray}
\partial_{\mu }{y^{k}}={h^{k}}_{\mu }\;,\\
\frac{\partial {h^{k}}_{\mu }}{\partial x^{\nu }}=
\mbox{\tiny{$\left\{\mbox{\hspace*{-1.5ex}}\begin{array}{c}\mbox
{\scriptsize $\lambda $}\\ \mbox{\scriptsize $\mu \: \nu $}
\end{array}\mbox{\hspace*{-1.5ex}}\right\}$}}{h^{k}}_{\lambda }\;,\\
{h^{k}}_{\mu }g^{\mu \nu }{h^{l}}_{\nu }=\eta^{kl}\;,
\end{eqnarray}
where $\mbox{\tiny{$\left\{\mbox{\hspace*{-1.5ex}}\begin{array}{c}\mbox
{\scriptsize $\lambda $}\\ \mbox{\scriptsize $\mu \: \nu $}
\end{array}\mbox{\hspace*{-1.5ex}}\right\}$}}$ stands for the
Christoffel symbol,
\begin{equation}
\mbox{\tiny{$\left\{\mbox{\hspace*{-1.5ex}}\begin{array}{c}\mbox
{\scriptsize $\lambda $}\\ \mbox{\scriptsize $\mu \: \nu $}
\end{array}\mbox{\hspace*{-1.5ex}}\right\}$}}
\stackrel{\rm def}{=}\frac{1}{2}g^{\lambda \xi }
(\partial_{\mu }g_{\xi \nu }+ \partial_{\nu }g_{\xi \mu }-
\partial_{\xi }g_{\mu \nu })\;.
\end{equation}
Let us examine (3.9)--(3.11). From (3.1), (3.9), and
(3.11), it follows that the Jacobian
$\det(\partial_{\mu }y^{k})$ has the expression
\begin{equation}
\det(\partial_{\mu }y^{k})
=\det({h^{k}}_{\mu })=\pm AB^{2}\;.
\end{equation}
For the solution (3.2), the right hand side of (3.13) diverges
for $r=0$ and vanishes for
$r=l$, which implies that the coordinate transformation between
$\{x^{\mu };\mu =0,1,2\}$ and any \lq \lq Minkowskian" coordinate
system $\{y^{k};k=0,1,2\}$ is singular not only at the origin $r=0$,
but also on the event horizon $r=l$.
Equation (3.10) takes the form,
\begin{eqnarray}
\left\{ \begin{array}{ll}\displaystyle \partial_{0}{h^{k}}_{0}=
  \frac{a}{r_{0}^{2}}Ax^{\alpha }{h^{k}}_{\alpha }\;, \\
  \displaystyle \partial_{0}{h^{k}}_{\alpha }=
  \partial_{\alpha }{h^{k}}_{0}=
  \frac{a}{Ar^{2}}x^{\alpha }{h^{k}}_{0}\;, \\
  \displaystyle \partial_{\alpha }{h^{k}}_{\beta }=
  \frac{1}{r^{2}}(\delta_{\alpha \beta }x^{\gamma }{h^{k}}_{\gamma }
  -x^{\alpha }{h^{k}}_{\beta }-x^{\beta }{h^{k}}_{\alpha })\;, \\
\mbox{\hspace{28.9ex} $\alpha \; , \; \beta = 1 \; , 2 \; ,$}
                                   \end{array}\right.
\end{eqnarray}
for this solution. From the first two of (3.14), we can derive an
equation for ${h^{k}}_{0}$, which is solved to give
\begin{equation}
{h^{k}}_{0}=I^{k}(x^{\alpha })\exp \left(\frac{ax^{0}}{r_{0}}\right)
+J^{k}(x^{\alpha })\exp \left(-\frac{ax^{0}}{r_{0}}\right)\;
\end{equation}
with $I^{k}$ and $J^{k}$ being functions of $x^{\alpha }$.
Substituting (3.15) into the second of (3.14) and into (3.9) with
$\mu =0$, and integrating the equations obtained, we get
\begin{eqnarray}
{h^{k}}_{\alpha }& = & \frac{r_{0}x^{\alpha }}{Ar^{2}}
\left[I^{k}(x^{\beta })\exp \left(\frac{ax^{0}}{r_{0}}\right)-
J^{k}(x^{\beta })\exp \left(-\frac{ax^{0}}{r_{0}}\right)\right]
\nonumber \\
& &+{X^{k}}_{\alpha }(x^{\beta })\;,
\end{eqnarray}
\begin{equation}
y^{k}=\frac{r_{0}}{a}
\left[I^{k}\exp \left(\frac{ax^{0}}{r_{0}}\right)
-J^{k}\exp \left(-\frac{ax^{0}}{r_{0}}\right)\right]
+Y^{k}(x^{\beta })
\end{equation}
\noindent with ${X^{k}}_{\alpha }$ and $Y^{k}$ being functions
of $x^{\beta }$.
{}From (3.9) with $\mu =\alpha $, (3.16), and (3.17), it follows that
\begin{eqnarray}
\left\{ \begin{array}{ll}\displaystyle
\partial_{\alpha }I^{k}=\frac{a}{Ar^{2}}x^{\alpha }I^{k}\;,\\
\displaystyle
\partial_{\alpha }J^{k}=\frac{a}{Ar^{2}}x^{\alpha }J^{k}\;,\\
\displaystyle
\partial_{\alpha }Y^{k}={X^{k}}_{\alpha }\;.
                                       \end{array}\right.
\end{eqnarray}

By substituting (3.16) into the last of (3.14), we obtain
\begin{equation}
\partial_{\alpha }\partial_{\beta }Z^{k}=\frac{1}{r^{2}}
[\delta_{\alpha \beta }x^{\gamma }\partial_{\gamma }Z^{k}
-x^{\alpha }\partial_{\beta }Z^{k}-x^{\beta }
\partial_{\alpha }Z^{k}]
\end{equation}
with $Z^{k}$ standing for any of $I^{k}, J^{k}$ and $Y^{k}$.
Equations (3.19) with $Z^{k}=I^{k}$ and (3.19) with
$Z^{k}=J^{k}$ are satisfied, if the first of (3.18) and the second
of (3.18) are satisfied, respectively. Equation (3.19) with
$Z^{k}=Y^{k}$ is solved to give
\begin{equation}
Y^{k}=A^{k}+B^{k}\ln r+C^{k}\theta
\end{equation}
with $A^{k}, B^{k}$, and $C^{k}$ being integration constants.
Here, we have used the coordinate $(r, \theta )\;;
x^{1}=r\cos \theta \;,\; x^{2}=r\sin \theta $.

Integrating the first and the second of (3.18), we obtain
\begin{equation}
I^{k} = D^{k}A\;, \; \; \; J^{k}=E^{k}A\;,
\end{equation}
where $D^{k}$ and $E^{k}$ are integration constants.

{}From (3.17), (3.20), and (3.21), it follows that
\begin{eqnarray}
y^{k} &=& \frac{r_{0}A}{a}
\left[D^{k}\exp \left(\frac{ax^{0}}{r_{0}}\right)
-E^{k}\exp \left(-\frac{ax^{0}}{r_{0}}\right)\right] \nonumber \\
& &+A^{k}+B^{k}\ln r+C^{k}\theta \;.
\end{eqnarray}

The constants $A^{k}, B^{k},C^{k}, D^{k}$, and $E^{k}$ have to be
chosen so as to satisfy (3.11). The following is easily known:
\begin{description}
\item[(A)]$A^{k}\;(k=0,1,2)$ are free parameters.
\item[(B)]The constants $B^{k}\;(k=0,1,2)$ should be chosen as
$(B^{1}, B^{2}, B^{3})=(0,0,0)$, so far as all of $B^{k}, C^{k}$,
and $D^{k}$ are required to be real. The choice
$(C^{1}, C^{2}, C^{3})=(0,0,0)$ is excluded,
which implies that at least one of the coordinates
$y^{k}\;(k=0,1,2)$ has a \lq \lq compact part" proportional to
$\theta , 0\leq \theta \leq 2\pi $.
\item[(C)]There are many sets of $(B^{k}=0,C^{k}, D^{k}, E^{k})$
satisfying (3.11), among which we have the following:
\begin{eqnarray}
\left\{ \begin{array}{ll}\displaystyle
C^{0}=C^{2}=D^{1}=E^{1}=0, C^{1}=r_{0}\;,\\
\displaystyle
D^{0}=E^{0}=\frac{1}{2}, D^{2}=-E^{2}=\frac{1}{2}\varepsilon \;, \\
\displaystyle
\mbox{\hspace{22.7ex} $\varepsilon \stackrel{\rm def}{=} \pm 1\;,$}
\end{array}\right.
\end{eqnarray}
which gives
\begin{eqnarray}
\left\{ \begin{array}{ll}\displaystyle
y^{0}=\frac{r_{0}}{a}
A\sinh \left(\frac{a}{r_{0}}x^{0}\right)+A^{0}\;,\\
\displaystyle
y^{1}=r_{0}\theta +A^{1}\;,\\
\displaystyle
y^{2}=\varepsilon \frac{r_{0}}{a}
A\cosh \left(\frac{a}{r_{0}}x^{0}\right)+A^{2}\;.
\end{array}\right.
\end{eqnarray}
{}From (3.24), the following is known:
\begin{description}
\item[(a)]The coordinate $y^{1}$ given by (3.24) takes values
in the closed interval $[A^{1}, A^{1}+2\pi r_{0}]$, and the point
$(y^{0}, y^{1}=A^{1}, y^{2})$ and the point $(y^{0},
y^{1}=A^{1}+2\pi r_{0}, y^{2})$ should be identified with each other.
\item[(b)]There is a relation
\begin{equation}
\eta_{kl}(y^{k}-A^{k})(y^{l}-A^{l})
={\left(\frac{r_{0}}{a}A\right)}^{2}+{(r_{0}\theta )}^{2}\;,
\end{equation}
the right hand side of which is non-negative and vanishes only for
$(r,\theta )=(l, 0)$. Thus, the space-time is embedded in the
non-time-like region $\eta_{kl}(y^{k}-A^{k})(y^{l}-A^{l})\geq 0$.
\end{description}
Also, from (3.9) and (3.24), we obtain
\widetext
\begin{equation}
({h^{k}}_{\mu })=\left(\begin{array}{ccc} \displaystyle
A\cosh \left(\frac{a}{r_{0}}x^{0}\right) &
\displaystyle
\frac{r_{0}x^{1}}{r^{2}}\sinh \left(\frac{a}{r_{0}}x^{0}\right) &
\displaystyle
\frac{r_{0}x^{2}}{r^{2}}\sinh \left(\frac{a}{r_{0}}x^{0}\right)\\
\displaystyle
0 &
\displaystyle
-\;\frac{r_{0}x^{2}}{r^{2}} &
\displaystyle
\frac{r_{0}x^{1}}{r^{2}}\\
\displaystyle
\varepsilon A\sinh \left(\frac{ax^{0}}{r_{0}}\right) &
\displaystyle
\varepsilon \;\frac{r_{0}x^{1}}{r^{2}}
\cosh \left(\frac{a}{r_{0}}x^{0}\right) &
\displaystyle
\varepsilon \;\frac{r_{0}x^{2}}{r^{2}}
\cosh \left(\frac{a}{r_{0}}x^{0}\right)\\
\end{array}\right)\;.
\end{equation}
\end{description}
\narrowtext
\vspace{5mm}

 Finally, we comment on the solution (3.3). This gives the metric
\begin{equation}
g_{00}=-1\;,\; \; \; g_{0\alpha }=0\;, \; \; \; g_{\alpha \beta }=
\left(\frac{r}{r_{0}}\right)^{2b}\delta_{\alpha \beta }\;,
\end{equation}
which agrees with the metric given by (2.7) of Ref. [5] for the case
with $n=1\;, r_{1}=0\;, m_{1}=m$, if and only if
\begin{equation}
{r_{0}}^{-2b}=C\;, \; \; \; b=-4Gm\;,
\end{equation}
and this solution does not give a black hole space-time.
\section{SUMMARY AND COMMENTS}
                                        \setcounter{equation}{0}
The above results can be summarized as follows:
\begin{description}
\item[1.]The solutions (3.2) and (3.3) of the gravitational field
equation in vacuum for the case with $3\alpha =-4\beta \neq 0$ are
also solutions of the three-dimensional vacuum
Einstein equation with a vanishing cosmological constant.
\item[2.]The solution (3.2) gives a Newtonian limit.
\item[3.]For the solution (3.2), the Riemann-Christoffel tensor
vanishes for $r\neq 0$, and it gives a black hole in a
\lq \lq flat-land" in the Einstein theory. The coordinate
transformation between $\{x^{\mu };\mu =0,1,2\}$ and any
\lq \lq Minkowskian" coordinate system $\{y^{k};k=0,1,2\}$,
however, is singular not only at the origin $r=0$, but also on
the event horizon $r=l$. Also, the space-time has a rich structure,
as is described by the statements (a) and (b) in the preceding
section. Even as a solution in the Einstein theory, (3.2) does not
give a space-time which is flat in the region $r\neq 0$ in the true
sense of the word.
\item[4.]The solution (3.2) gives a black hole space-time also in the
teleparallel theory. This space-time is not flat in any rude sense,
because the torsion tensor does not vanish there.
\item[5.]The solution (3.3) gives a metric which essentially agrees
with a metric given by Deser, Jackiw, and 't Hooft [5].
\end{description}

The following should be mentioned here:
\begin{description}
\item[a.]As we have shown in Ref. [1], our theory has a Newtonian
limit, if the condition (5.11) of Ref. [1] is satisfied.
The condition $3\alpha =-4\beta \neq 0$, on which the solution (3.2)
is obtained, violates (5.11) of Ref. [1]. Neverthless, (3.2) gives
a Newtonian limit. But, this is not a contradiction, because
(3.2) does not satisfy, for $r \simeq r_{0}$, the condition
$A^{\prime } \simeq 0 \simeq B^{\prime }$ on which
(5.11) of Ref. [1] is based.
\item[b.]Since the solution (3.2) is also a solution of the
three-dimensional vacuum Einstein equation, {\em the three-dimensional
Einstein theory can have a Newtonian limit}, which is opposed to the
commonly-accepted claim [6-8] that this theory does not have the
limit.
\end{description}

The solution (3.2) gives a space-time which is exotic in several
respects, and vacuum regions of space-times in the three-dimensional
Einstein theory can be locally non-trivial.


\begin{references}
\bibitem[2]{spin}By \lq \lq spin" we mean the quantum number
 associated with the three-dimensional Lorentz group.
\bibitem[4]{curv}The Riemann-Christoffel curvature tensor should not
 be confused with ${R^{\mu }}_{\nu \lambda \rho }$ defined by the
 first part of (2.6).
\end{references}
\end{document}